\documentclass[conference]{IEEEtran}
\makeatletter
\def\ps@IEEEtitlepagestyle{%
	\def\@oddfoot{\mycopyrightnotice}%
	\def\@evenfoot{}%
}
\def\mycopyrightnotice{%
	{ }% <--- Change here
	\gdef\mycopyrightnotice{}% just in case
}

\usepackage{cite}
\ifCLASSINFOpdf
\else
\fi

\usepackage{times}
\usepackage{graphicx} 
\usepackage{subfigure} 
\usepackage{amsmath} 
\usepackage{amssymb}

\usepackage{algorithm}
\usepackage{algorithmic}

\usepackage{kantlipsum}
\usepackage{float}
\begin{document}
\title{A New Non-parametric Process Capability Index}

\author{\IEEEauthorblockN{Deovrat Kakde}
\IEEEauthorblockA{SAS Institute Inc.\\
SAS Campus Dr.,\\
Cary, NC 27513, USA\\
Email: dev.kakde@sas.com}
\and
\IEEEauthorblockN{Arin Chaudhuri}
\IEEEauthorblockA{SAS Institute Inc.\\
SAS Campus Dr.,\\
Cary, NC 27513, USA\\
Email: arin.chaudhuri@sas.com}
\and
\IEEEauthorblockN{Diana Shaw}
\IEEEauthorblockA{SAS Institute Inc.\\
SAS Campus Dr.,\\
Cary, NC 27513, USA\\
Email: diana.shaw@sas.com}}


%Telephone: (800) 555--1212, Fax: (888) 555--1212}


\maketitle

\begin{abstract}
%\boldmath
Process capability index (PCI) is a commonly used statistic to measure ability
of a process to operate within the given specifications or to produce 
products which meet the required quality specifications. PCI can be univariate or 
multivariate depending upon the number of process specifications or quality characteristics
of interest. Most PCIs make distributional assumptions which are often unrealistic in practice.

This paper proposes a new multivariate non-parametric process capability index. This index can be used when 
distribution of the process or quality parameters is either unknown or does not follow commonly used distributions 
such as multivariate normal.

\end{abstract}


% \fi
\IEEEpeerreviewmaketitle

\section{Introduction}

A process capability index (PCI) is an objective measure of the ability of a manufacturing process to produce products that conform to the design specifications. PCI is an important tool in statistical process control. PCI is used to quantify the relationship between process performance and process specification limits \cite{wu2009overview}.  The process capability index is used as a criterion for qualifying production machinery for specific products. Manufacturers often demand process capability indexes from their suppliers as evidence of a supplier's quality assurance practices. \\
The simplest and most commonly used univariate PCIs are $\mathrm{C_p}$ and $\mathrm{C_{pk}}$. Consider a manufacturing process, with one process parameter of interest, $\mathrm{x}$. Let $\mathrm{USL}$ and $\mathrm{LSL}$ denote the upper and lower engineering specification limits for $\mathrm{x}$. Specification limits define the acceptable values of a process or a quality parameter. Assuming that $\mathrm{\mu}$ and $\mathrm{\sigma}$ denote the mean and the standard deviation of $\mathrm{x}$, $\mathrm{C_p}$ and $\mathrm{C_{pk}}$ are defined as:\\

$\mathrm{C_p = \dfrac{USL-LSL}{6\sigma} }$\\

and 

$\mathrm{C_{pk} = \min({\dfrac{USL-\mu}{3\sigma}, \dfrac{\mu-LSL}{3\sigma} })} $ \\

The PCI $\mathrm{C_p}$ is computed as a ratio of tolerance band (defined as the interval between $\mathrm{USL}$ and $\mathrm{LSL}$) to the process variability. It assumes process parameter of interest follows a normal distribution with the process mean centered at the midpoint of the tolerance band. High $\mathrm{C_p}$ value, greater than one is desired. Such value indicates that process variability is low as compared to the tolerance band and it indicates ability of the process to meet specifications.\\
Often manufacturing processes are not centered at the midpoint of tolerance band. For such processes, PCI $\mathrm{C_{pk}}$ is recommended. For a process centered at tolerance midpoint value, $\mathrm{C_p}$ and $\mathrm{C_{pk}}$ are identical.

Many manufacturing processes have two or more correlated parameters of interest, which need to be monitored and controlled. For such processes, multivariate PCI are used.  An excellent overview of various multivariate process capability indexes is provided in \cite{zahid2008assessment}. One of the popular parametric multivariate PCI is the multivariate capability vector (abbreviated as MPCI) \cite{shahriari2009new,wang2000comparison}. This vector has three components. The first two components are computed with an assumption that the process data follows a multivariate normal distribution. The third component is based on a geometric understanding of
the process relative to the engineering specifications. It is expressed as $\mathrm{[C_{p}M, PV, LI]}$ where:
\begin{enumerate}
	\setcounter{enumii}{4}
	\item $\mathrm{C_{p}M}$ is computed using ratio of volume enclosed by the specification limits to the volume enclosed by the process spread. It is similar to the univariate PCI $\mathrm{C_p}$.
	\item $\mathrm{PV}$ is the probability value indicative of distance of the process mean from the tolerance midpoint. Higher value, close to 1 indicates that the process center is near tolerance midpoint. 
	\item $\mathrm{LI}$ is the indicator variable which takes a value of 1 if spread of the process is contained within the specification limits, 0 otherwise.
\end{enumerate}

In this paper, we propose a new Support Vector Data Description (SVDD) based process capability vector, $\mathrm{PC_{SVDD}}$. The proposed process capability vector makes no distributional assumptions. The components of the $\mathrm{PC_{SVDD}}$ vector are similar to those proposed by \cite{shahriari2009new}. 

The rest of the paper is organized as follows. Section II provides a brief introduction to SVDD. Computational details of $\mathrm{PC_{SVDD}}$ are provided in section III, followed by some examples in section IV. A comparison of MPCI with $\mathrm{PC_{SVDD}}$ is provided in section V  and conclusions are provided in section VI.
 
\section{Support Vector Data Description} 

Support Vector Data Description (SVDD) is a machine learning technique useful for one class classification and outlier detection. SVDD technique is similar to Support Vector Machines and was first introduced by Tax and Duin in \cite{tax2004support}. SVDD is used in domains where the majority of data belongs to a single class, or when one of the classes is significantly undersampled. The SVDD algorithm builds a flexible boundary around the target class data; this data boundary is characterized by observations designated as support vectors. Applications of SVDD include machine condition monitoring \cite{widodo2007support, ypma1999robust}, image classification \cite{sanchez2007one}, and multivariate process control \cite{sukchotrat2009one, kakde2017non}.

SVDD has the advantage that no assumptions about the distribution of the data need to be made. The technique can describe the shape of the target class without prior knowledge of the specific data distribution, with observations falling outside of the data boundary flagged as potential outliers. 

The $\mathrm{PC_{SVDD}}$ proposed in this paper exploits the ability of SVDD to correctly determine the boundary of any arbitrary shaped data.  

\subsection{Mathematical Formulation of SVDD}
\label{mfsvdd}
{\bf Normal Data Description}\\
The SVDD model for normal data description builds a minimum-radius hypersphere around the data.\\ 
\\
{\bf Primal Form}\\
Objective function:
\begin{equation}
\min R^{2} + C\sum_{i=1}^{n}\xi _{i} 
\end{equation}
subject to: 
\begin{align}
\|x _{i}-a\|^2 \leq R^{2} + \xi_{i}, \forall i=1,\ldots,n\\
\xi _{i}\geq 0, \forall i=1,\ldots,n
\end{align}
where:\\
$x_{i} \in {\mathbb{R}}^{m}, i=1,\ldots,n  $ represents the training data,\\
$ R$ is the radius and represents the decision variable,\\
$\xi_{i}$ is the slack for each variable,\\
$a$ is the center, \\
$C=\frac{1}{nf}$ is the penalty constant that controls the tradeoff between the volume and the errors, and\\
$f$ is the expected outlier fraction.\\ \ \\
{\bf Dual Form}\\
The dual formulation is obtained using the Lagrange multipliers.\\ 
Objective function:
\begin{equation} 
\max\ \sum_{i=1}^{n}\alpha _{i}(x_{i}\cdot x_{i}) - \sum_{i,j}^{ }\alpha _{i}\alpha _{j}(x_{i} \cdot x_{j}) 
\end{equation}
subject to:
\begin{align}
&   \sum_{i=1}^{n}\alpha _{i}  = 1\\
&  0 \leq  \alpha_{i}\leq C,\forall i=1,\ldots,n
\end{align}
where\\
$\alpha_{i}\in \mathbb{R}$ are the Lagrange constants and\\
$C=\frac{1}{nf}$ is the penalty constant.\\ \ \\
{\bf Duality Information}\\
Depending upon the position of the observation, the following results hold:\\
Center position: \begin{equation} \sum_{i=1}^{n}\alpha _{i}x_{i}=a \end{equation}
Inside position: \begin{equation} \left \| x_{i}-a \right \| < R \rightarrow \alpha _{i}=0 \end{equation}
Boundary position: \begin{equation} \label{eq:9} \left \| x_{i}-a \right \| = R \rightarrow 0< \alpha _{i}< C\end{equation}
Outside position: \begin{equation} \label{eq:10} \left \| x_{i}-a \right \| > R \rightarrow \alpha _{i}= C\end{equation}
The circular data boundary can include a significant amount of space in which training observations are very sparsely distributed. Scoring with this model can increase the probability of false positives.
Hence, instead of a circular shape, a compact bounded outline around the data is often desired. Such an outline should approximate the shape of the single-class training data. This is possible with the use of kernel functions.\\
{\begin{flushleft}
		\bf {Flexible Data Description}\\	
	\end{flushleft}
	The support vector data description is made flexible by replacing the inner product $ (x_{i}\cdot x_{j}) $ with a suitable kernel function $ K(x_{i},x_{j}) $. The Gaussian kernel function used in this paper is defined as
	\begin{equation}  
	K(x_{i}, x_{j})= \exp  \dfrac{ -\|x_i - x_j\|^2}{2s^2}
	\end{equation}
	where $s$ is the Gaussian bandwidth parameter.
	Results 7 through 10 hold when the kernel function is used in the mathematical formulation.\\
	The threshold $R^{2}$ is calculated as
	\begin{equation}
	R^{2} = K(x_{k},x_{k})-2\sum_{i}^{ }\alpha _{i}K(x_{i},x_{k})+\sum_{i,j}^{ }\alpha _{i}\alpha _{j}K(x_{i},x_{j})
	\end{equation}
	using any $ x_{k} \in SV_{<C} $
	, where $SV_{<C}$  is the set of support vectors for which $ \alpha _{k} < C $.
	\begin{flushleft}
		{\bf Scoring}	
	\end{flushleft}
	For each observation $ z $  in the scoring data set, the distance $ dist^{2}(z) $ is calculated as follows: 
	\begin{equation} dist^{2}(z)= K(z,z) - 2\sum_{i}^{ }\alpha _{i}K(x_{i},z) +\sum_{i,j}^{ }\alpha _{i}\alpha _{j}K(x_{i},x_{j})\end{equation}
	Observations in the scoring data set for which $ dist^{2}(z) > R^{2} $ are designated as outliers.

\subsection{Importance of kernel bandwidth value}
The flexible data description is preferred when data boundary
needs to closely follow the shape of data. The tightness of the boundary is a function of
the number of support vectors. In the case of a Gaussian kernel, it is observed
that if the value of the outlier fraction $f$ is kept constant, the number of
support vectors identified by the SVDD algorithm is a function of the Gaussian
bandwidth $s$. At a very low value of $s$, the number of support vectors is
large and approaching the number of observations. As the value of $s$
increases, the number of support vectors is reduced. It is also observed
that at lower values of $s$ the data boundary is extremely wiggly.
As $s$ increases, the data boundary becomes less wiggly
and it starts to follow the shape of the data. There are several methods for setting an
appropriate kernel bandwidth value. Some of the unsupervised methods include the Peak criterion \cite{kakde2017peak,pered8258344} and the Mean criterion \cite{chau8215749}. These methods can be used to select an appropriate value of the Gaussian bandwidth value which can provide a geometric boundary which takes essential geometric features of the data into consideration. The references \cite{kakde2017peak,pered8258344, chau8215749} provide multiple examples where the selected value of bandwidth parameter has provided accurate description of training data in a geometric sense.

\section{SVDD Based Process Capability Index}

As mentioned in section I, process capability index is computed by comparing the spread of a process under a state of statistical control to the corresponding engineering specification limits. This section details the  SVDD-based process capability vector, $\mathrm{PC_{SVDD}}$.  This vector is defined as\\
\begin{equation}
\mathrm{PC_{SVDD}}= \mathrm{[C_p, dist, p]}\\
\end{equation}
where\\
\\
$\mathrm{C_p}$ = $\displaystyle\frac{\mbox{Volume of engineering specification limits}}{\mbox{Volume of process region}}$,\\

$\mathrm{dist}$ is the Euclidean distance between the process center $\mathrm{a}$ and the center of engineering specification, and\\

$\mathrm{p}$ is the fraction of process observations that fall outside the engineering specifications. \\

Pattnaik and Tripathy \cite{patt6962567} proposed a SVDD based process capability index $\mathrm{MC_{SVDD}}$ which uses computational geometry to compute the convex hull and volume of the process data. The $\mathrm{PC_{SVDD}}$ proposed in this paper uses a simulation based approach with SVDD scoring, which is computationally very efficient as compared to \cite{patt6962567}.

\subsection{{$ \mathrm{PC_{SVDD}}$} Computation}
The following sections provide guidelines for computing components of the $\mathrm{PC_{SVDD}}$ vector.
\subsubsection{$\mathrm{C_p}$ Computation}
This paper proposes a simulation-based approach for computing $\mathrm{C_p}$. A higher value of $\mathrm{C_p}$ is desirable and indicates that a process spread is very narrow compared to the engineering specifications. The $\mathrm{C_p}$ computations require following inputs:
\begin{enumerate}
	\item Number of variables $q$.
	\item Upper and lower engineering specification limits  [$\text{USL}_j$, $\text{LSL}_j$] for $j$ = 1 to $q$
	\item Number of observations to be simulated using engineering specifications $\mathrm{N_{ES}}$
	\item Process measurement window $\mathrm{W}$
	\item Gaussian bandwidth parameter $s$
	\item Fraction outlier $f$. In the following computations, the value of $f$ can be set to a very low value such as $\mathrm{1e-6}$. The rationale behind this is that any process capability analysis requires process to be in a state of statistical control. Such state does not have any assignable or special causes of variation, hence will generate measurements with very low fraction outlier $f$.	
\end{enumerate}

Steps for calculating $\mathrm{C_p}$ are as follows:\\

\textbf{Step 1:} Obtain the set of support vectors $SV$, threshold value $R^{2}$, set of Lagrange coefficients $\alpha$, and center $a$ by training SVDD on window $W$ to perform SVDD computations.\\
$<SV, R^{2},\alpha, a> \leftarrow \text{Train}_{SVDD}(W) $\\

\textbf{Step 2:} Simulate data set $S$, which contains $\mathrm{N_{ES}}$ observations that are uniformly distributed between [$\text{USL}_j$, $\text{LSL}_j$] for $j$ = 1 to $q$.  \\

\textbf{Step 3:} Score each observation $i$ in data set $S$ by using $<SV,\alpha, R^{2}>$. Obtain $dist^2(i)$ for each observation $i$, for $i$ = 1 to $\mathrm{N_{ES}}$  \\

\textbf{Step 4:} Obtain $\text{COUNT}_1$, the count of observations for which $dist^2(i) \leq R^2$.\\

\textbf{Step 5:} Compute $C_p$ as\\

\begin{flalign}
C_p &= \displaystyle\frac{\mbox{Volume of engineering specification limits}}{\mbox{Volume of process region}} \\ 
&= \displaystyle\frac{\mbox{$\mathrm{N_{ES}}$}}{\mbox{$\mathrm{COUNT_1}$}}.	 \\
\end{flalign}          

\subsubsection{$\mathrm {dist}$ Computation}
The $dist$ value refers to the euclidean distance between the process center $a$ and the center of engineering specifications $c$. The coordinates of the center $c$ are the midpoints of the corresponding upper and lower specification limits. $a$ is computes using equation [7]. \\
\begin{equation}
\mathrm{dist} = \lVert \mathbf{a-c} \rVert
\end{equation}
\subsubsection{$\mathrm{p}$ Computation}
The $\mathrm{p}$ value represents the fraction of process observations that fall outside the engineering specifications. This value can be easily computed using a simple SQL query.
\section{Examples}
This section illustrates two examples of the $\mathrm{PC_{SVDD}}$ vector. The first example uses circular bivariate process data to explain the concept.  The second example uses steel sleeve production process data. Computations were performed using the SVDD procedure available in SAS$^{\tiny{\textregistered}}$ software \cite{institute2017sas}.\\
\subsubsection{Example 1}
The purpose of this example is to geometrically explain various components of the $\mathrm{PC_{SVDD}}$ vector. This example uses two process variables $\mathrm{x_1}$ and $\mathrm{x_2}$. Figure \ref{fig:Fig_21} (a) illustrates the process spread. Figure \ref{fig:Fig_21} (b) through (f) show the original data with engineering specification limits (rectangle surrounding the circle) and varying distance between the centers of process data and the engineering specification limits. Figure \ref{fig:Fig_21} (b) and (c) indicate scenarios in which the process center matches the center of the engineering specifications and the entire process data are contained within the specification limits. The scenario shown in Figure \ref{fig:Fig_21} (c) is desirable compared to the scenario in Figure \ref{fig:Fig_21} (b) because the $C_p$ (first component of the capability vector) is larger. Figure \ref{fig:Fig_21} (d) has a process mean that is shifted towards the right, so it has a $dist$ value of 1.  Figures \ref{fig:Fig_21} (e) and (f) have both a shifted mean and a  significant portion of the observations that lie outside the specification limits. This is reflected by nonzero values of $dist$ and $p$ in the corresponding process capability vectors. 

\begin{figure}{H}
	\begin{tabular}{ccc}
		\includegraphics[width=1.5in]{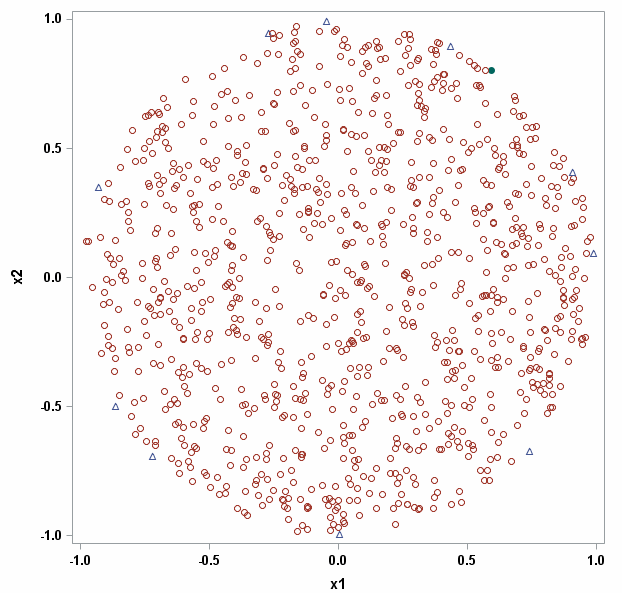} &   \includegraphics[width=1.5in]{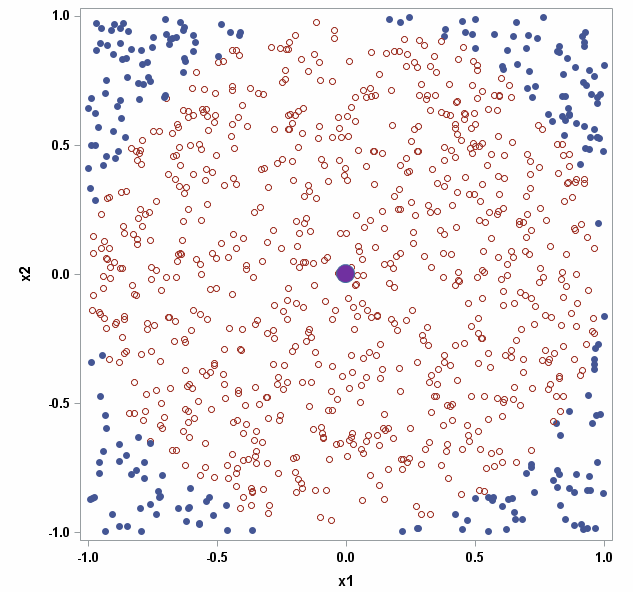} \\
		(a) Data & (b) [1.33, 0, 0] \\[5pt]
		\includegraphics[width=1.5in]{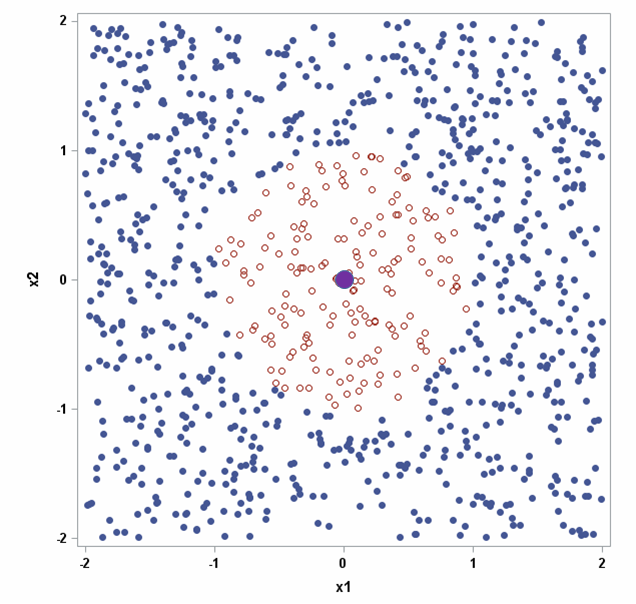} &   \includegraphics[width=1.5in]{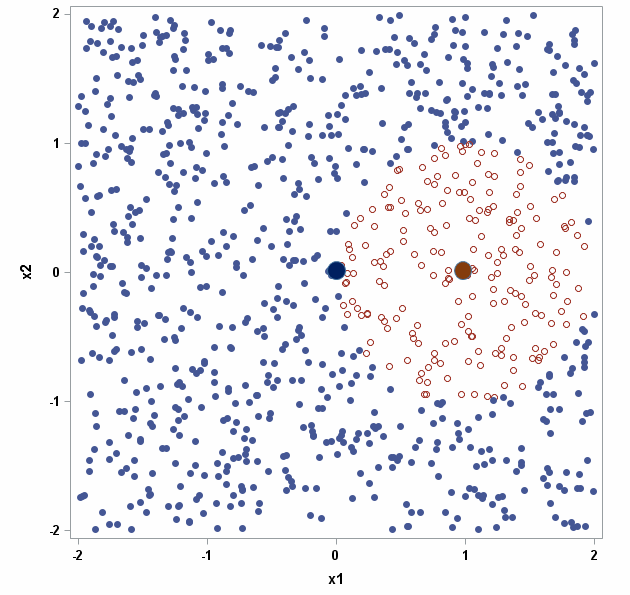} \\
		(c) [5.26, 0, 0] & (d) [5.84, 1, 0]\\[5pt]
		\includegraphics[width=1.5in]{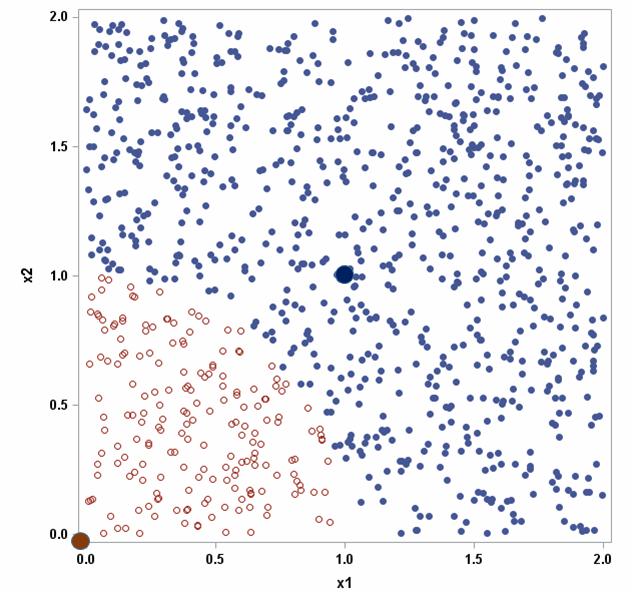} &   \includegraphics[width=1.5in]{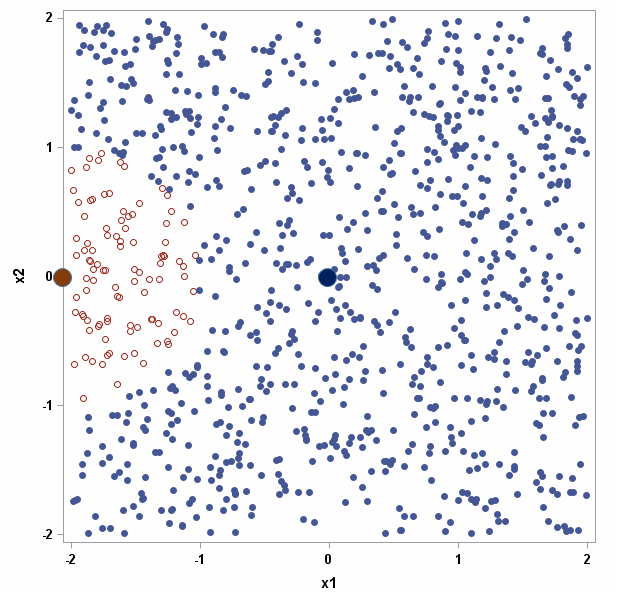} \\
		(e) [5.12, $\sqrt{2}$, 0.75] & (f) [10.1, 2, 0.50] \\[5pt]
	\end{tabular}
	\caption{ Results of Example 1. 	
		The brown unfilled markers indicate process observations, the smaller blue filled  markers indicate points identified as outliers that fall under the engineering specification limits, and the large markers indicate corresponding centers. 
	}\label{fig:Fig_21}
\end{figure}

\subsubsection{Example 2}
This example computes the $\mathrm{PC_{SVDD}}$ vector for the steel sleeve production process that is outlined in \cite{raissi2009}. The data set contains 28 observations, each of which consists of three diameter measurements (A, B, and C) at different parts of the sleeve.  The measurements were obtained from a process in a state of statistical control. Table \ref{table:t12} summarizes the parameters that are used for computing the $\mathrm{PC_{SVDD}}$ vector.
\begin{table}[]
	\centering
	\caption{Steel Sleeve Process Capability Example }
	\label{table:t12}
	\begin{tabular}{|l|l|}
		\hline
		\textbf{Parameter}  &  \textbf{Value}  \\ [0.5ex] 
		\hline	
		Number of variables &	3 \\
		\hline
		Number of observations  &	28 \\
		\hline
		Engineering specification limits: &	\\
		A &[171, 64] \\
		B &	[132, 0] \\
		C &	[147, 70] \\
		\hline
		Simulation data set size, $N_{ES}$ &	1,000,000 \\
		\hline
		Gaussian bandwidth parameter $s$  &	13 \\
		(computed using peak criteria \cite{kakde2017peak})&\\
		\hline
		Outlier fraction & 0.001\\		
		\hline
	\end{tabular}
\end{table}
The value of $\mathrm{PC_{SVDD}}$ is obtained as [43.2, 4.53, 0]. The value of 45.8 for the first component, $\mathrm{C_p}$, indicates that the process spread is very narrow compared to the engineering specifications. The $\mathrm{dist}$ value of 4.53 indicates that the process center is away from the center of the engineering specifications. The $\mathrm{p}$ value of 0 indicates that the process spread does not cut across the engineering specification limits. Figure \ref{fig:Fig_200} compares the process region to the engineering specifications.
\begin{figure}[]
	\centering
	\includegraphics[width=2.5in]{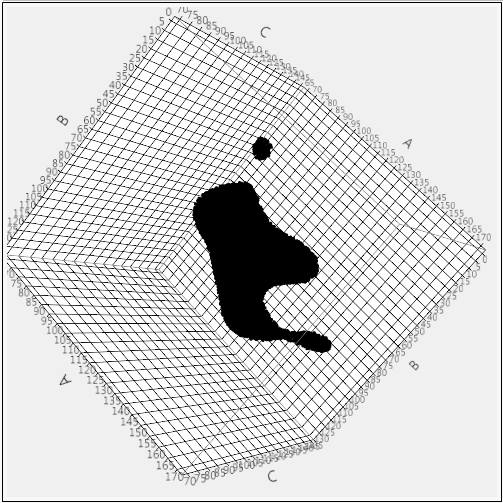}
	\caption{Steel Sleeve Process Spread}\label{fig:Fig_200}
\end{figure} 

\section{Comparison of MPCI and $\mathrm{PC_{SVDD}}$}

As outlined in section I, the multivariate process capability index (MPCI) assumes that the process parameters or the quality characteristics are distributed as multivariate normal. This translates into assuming that the shape of the underlying data cloud is ellipsoidal. This is a strong assumption. The distribution of process data is often unknown or non-normal \cite{das2009new, chakraborti2001nonparametric}. Many machines operate in multiple known operating modes. The process parameter data from such machines is generally multimodal. In some cases the number of operating modes in the machine is known. Some modern  machines based on their operating conditions, have ability to self adjust and move from one operating mode to another. The data from such machines is also multimodal, but number of operating modes is unknown. In such cases, the transition from one state to another may not be transparent when process data is being collected.The MPCI fits a single multivariate distribution to the collected data, which can provide misleading results when data is multimodal.\\
This section provides two examples which compare performance of MPCI with $\mathrm{PC_{SVDD}}$. These examples use bivariate synthetic data sets with two variables, x and y.  In the first example, the process data is boomerang shaped and in the second example the data is bimodal and consists of two donut shaped clusters. The USL, LSL and center for these two data sets are provided in Table \ref{table:t13}. Figure \ref{fig:Fig_300} (a) and (b) provide scatter plots of these two data sets, with USL and LSL for variables $\mathrm{x}$ and $\mathrm{y}$. The scatter plots indicate that the spread of the process is contained within the process specifications. Figure \ref{fig:Fig_300} (c) and (d) show the results for MPCI. These results were obtained using the MPCI package available in R \cite{santos2012mpci}. In Figure \ref{fig:Fig_300} (c) and (d), the red line indicates the engineering tolerance region, the dotted blue line shows the modified process region. The ellipse shows the fitted bivariate normal distribution. Figure \ref{fig:Fig_300} (e) and (f) show results for $\mathrm{PC_{SVDD}}$. These figures show the results obtained by scoring the $\mathrm{N_{ES}}$ observations in data set $S$ (refer to steps 2 and 3 in section II for $\mathrm{N_{ES}}$ and $S$). In Figure \ref{fig:Fig_300} (e) and (f), the gray area indicates observations classified as outliers and black area indicates observations classified as inliers. Comparison of Figure \ref{fig:Fig_300} (e) and (f) with Figure \ref{fig:Fig_300} (a) and (b) indicate that SVDD successfully captured the essential geometric features of the data. \\
\begin{table}[]
	\centering
	\caption{Specification Limits}
	\label{table:t13}
	\begin{tabular}{|l|l|l|}
		\hline
		& Boomerang  & Two-Donut  \\ \hline
	$\mathrm{USL_{x}}$	&12  & 20 \\ \hline
	$\mathrm{LSL_{x}}$	&-2  & -10 \\ \hline
	$\mathrm{USL_{y}}$	&12  & 30 \\ \hline
	$\mathrm{LSL_{y}}$	&-2 & -10 \\ \hline
	Center	& (5,5) & (5, 10)  \\ \hline
	\end{tabular}
	\end{table}
\begin{figure}[]
	\begin{tabular}{ccc}
		\includegraphics[width=1.5in]{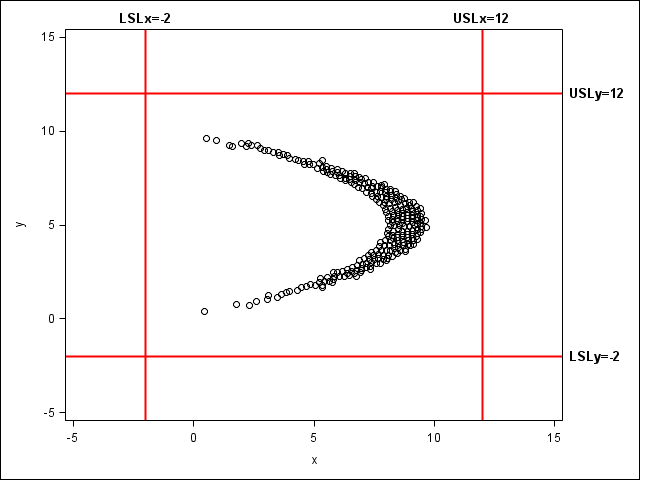} &   \includegraphics[width=1.5in]{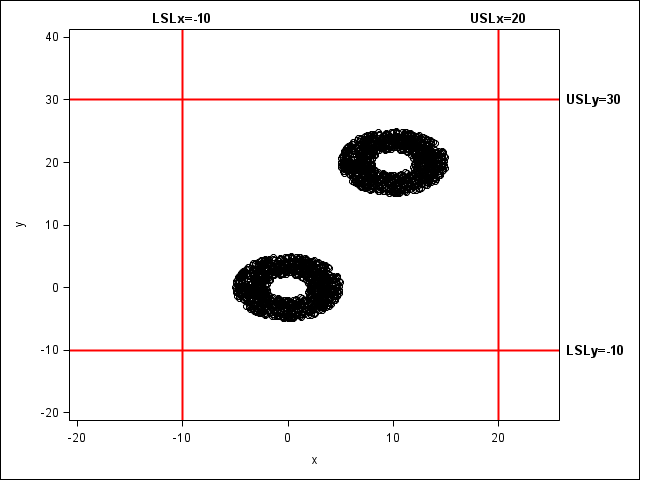} \\
		(a)  &  (b)  \\[5pt]\\
	   \includegraphics[width=1.5in]{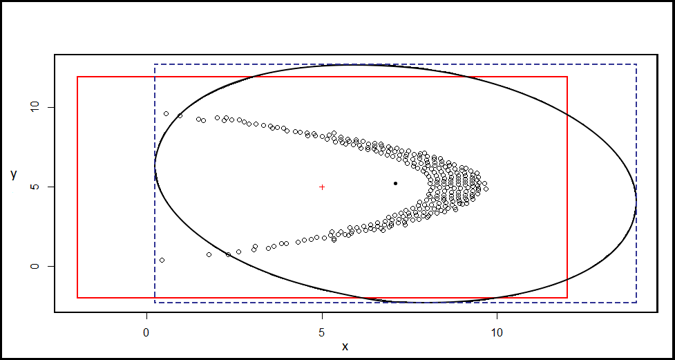} &   \includegraphics[width=1.5in]{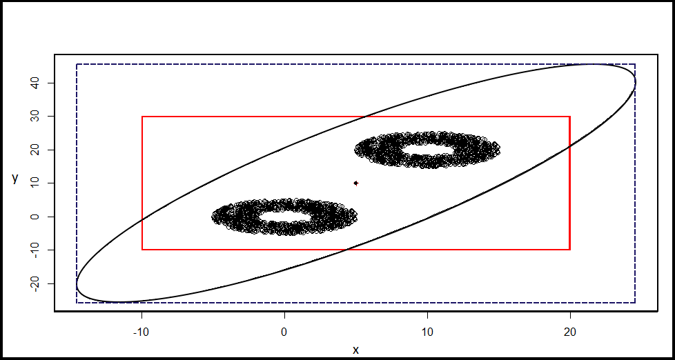} \\
	    (c)  &  (d)  \\[5pt]
 	   \includegraphics[width=1.5in]{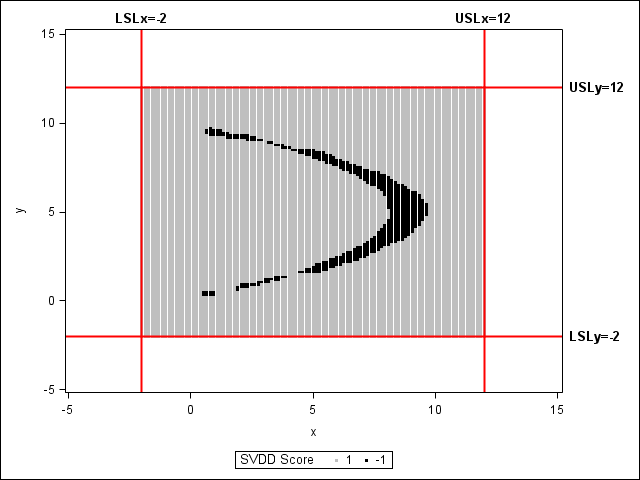} &   \includegraphics[width=1.5in]{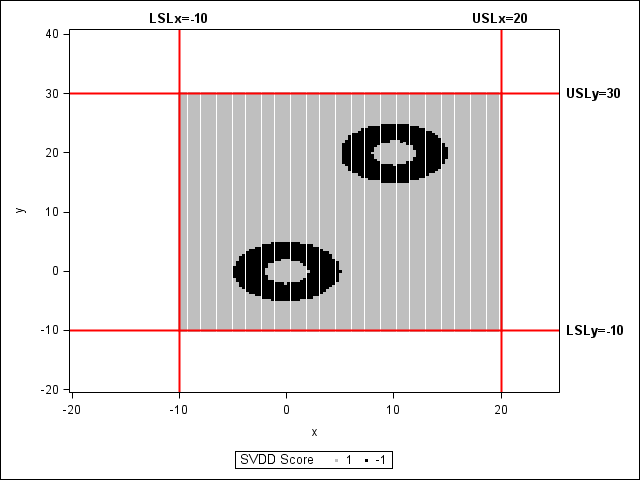} \\
	    (e)  &  (f)  \\[5pt]
	\end{tabular}
	\caption{ Comparison of MPCI and $\mathrm{PC_{SVDD}}$. 	
		Process Data using Boomerang and Two Donut data sets. Figures (a) and (b) show the process data with control limits. Figures (c) and (d) show the results MPCI results.
		Figures (e) and (f) show results for $\mathrm{PC_{SVDD}}$.
	}\label{fig:Fig_300}
\end{figure}
Table \ref{table:t15} shows the MPCI and $\mathrm{PC_{SVDD}}$ values for boomerang and two donut data sets. The $\mathrm{C_pM}$ parameter in MPCI and $\mathrm{C_p}$ parameter in $\mathrm{PC_{SVDD}}$ both relate to the ratio of volume enclosed by engineering specification limits to the volume of the process region. A ratio higher than one is desired. Higher ratio indicates that process is highly capable of operating within the specifications. The $\mathrm{C_p}$ parameter of $\mathrm{PC_{SVDD}}$ for boomerang and two donut data set is 18.996 and 9.307 respectively. Comparing these numbers to Figure \ref{fig:Fig_300} (a) and (b), indicates that $\mathrm{PC_{SVDD}}$ has correctly captured the ratio of volume enclosed by engineering specification limits to the volume of the process region. Where as the $\mathrm{C_pM}$ parameters of MPCI were 0.974 and 0.656, both indicating that under the assumption of multivariate normal distribution, the process is not capable of meeting the engineering specifications. \\
The second parameter $\mathrm{PV}$ of MPCI vector is the probability value indicative of distance of process mean from the tolerance midpoint. Higher value closer to 1 indicates process center is near the tolerance midpoint. The $\mathrm{PV}$ value is 0 for the boomerang data set  and 0.482 for the two donut data set. These numbers indicate that process center is more away from the tolerance midpoint for the boomerang data set than the two donut data set. In case of $\mathrm{PC_{SVDD}}$, the second parameter $\mathrm{dist}$ indicates the euclidean distance between the process center and tolerance mid-point, For both data sets, these values were very low.\\
The third parameter, $\mathrm{LI}$ of MPCI vector takes a value of 1 if the spread of the process is contained within the specification, 0 otherwise. The results show that the value of parameter $\mathrm{LI}$ is 0 for both data sets, which means under assumption of bivariate normal distribution, the spread of the process is not contained within the specification. The third parameter $\mathrm{p}$ of $\mathrm{PC_{SVDD}}$ represents the fraction of process observations that fall outside the engineering specifications. This parameter is 0 for both data sets, indicating that all process observations are contained within the engineering specification limits. \\

\begin{table}[H]
	\centering
	\caption{$\mathrm{PC_{SVDD}}$ Parameters}
	\label{table:t14}
	\begin{tabular}{|l|l|l|}
		\hline
		& Boomerang & Two Donut \\ \hline
	Simulation data set size, $\mathrm{N_{ES}}$	&10,201  &10,201  \\ \hline
		Gaussian bandwidth parameter $s$ 	& 0.7263714897  & 2.8127912992  \\ 
		(Computed using the mean criterion \cite{chau8215749}) &  & \\ \hline
		Outlier fraction&1E-6  &1E-6  \\ \hline
	\end{tabular}
\end{table}

\begin{table}[]
	\centering
	\caption{MPCI and $\mathrm{PC_{SVDD}}$ Values }
	\label{table:t15}
	\begin{tabular}{|l|l|l|}
		\hline
	    & Boomerang & Two Donut \\ \hline
	MPCI &  &  \\ 
	$[C_{p}M, PV, LI]$	& [0.974, 0, 0] & [0.656, 0.482, 0]  \\ \hline
	$\mathrm{PC_{SVDD}}$&  &  \\ 
	$[C_p, dist, p]$	& [18.996, 0.238, 0] & [9.307, 0.013, 0]  \\ \hline
	\end{tabular}
\end{table}

\section{Conclusions} 
\label{cn}
A new multivariate non-parametric process capability index $\mathrm{PC_{SVDD}}$ is presented in this paper. The process capability index is a vector with three entries $\mathrm{C_p, dist}$ and $\mathrm{p}$, which together provide assurance on ability of a process to conform to process specifications or produce products with required quality characteristics. The main contribution in this paper is use of SVDD to accurately compute $\mathrm{C_p}$, which provides a measure of ratio of volume of engineering specification to the volume of process spread.

\bibliographystyle{plain}
\bibliography{svdd_pci}
%\bibliographystyle{icml2016}

\end{document}